\documentclass[english,reprint, superscriptaddress, secnumarabic, nofootinbib, nobibnotes]{revtex4-1}
\usepackage[T1]{fontenc}
\usepackage[latin9]{inputenc}
\usepackage{geometry}
\usepackage{amsmath}
\usepackage{amsthm}

\makeatletter

\providecommand{\tabularnewline}{\\}

\newcommand{\R}{\mathcal{R}}

\newcommand{\C}{\mathsf{C}}

\def\frontmatter@abstractheading{}

\makeatletter
\renewcommand{\p@subsection}{}
\renewcommand{\p@subsubsection}{}
\makeatother

\makeatother

\usepackage{babel}
\begin{document}

\title{Replacing Nothing with Something Special:\\Contextuality-by-Default
and Dummy Measurements}

\author{Ehtibar N. Dzhafarov}
\email{E-mail: ehtibar@purdue.edu}

\selectlanguage{english}%

\affiliation{Purdue University, USA}
\begin{abstract}
The object of contextuality analysis is a set of random variables
each of which is uniquely labeled by a content and a context. In the
measurement terminology, the content is that which the random variable
measures, whereas the context describes the conditions under which
this content is measured (in particular, the set of other contents
being measured ``together'' with this one). Such a set of random
variables is deemed noncontextual or contextual depending on whether
the distributions of the context-sharing random variables are or are
not compatible with certain distributions imposed on the content-sharing
random variables. In the traditional approaches, contextuality is
either restricted to only consistently-connected systems (those in
which any two content-sharing random variables have the same distribution)
or else all inconsistently-connected systems (those not having this
property) are considered contextual. In the Contextuality-by-Default
theory, an inconsistently connected system may or may not be contextual.
There are several arguments for this understanding of contextuality,
and this note adds one more. It is related to the fact that generally
not each content is measured in each context, so there are ``empty''
content-context pairs. It is convenient to treat each of these empty
pairs as containing a dummy random variable, one that does not change
the degree of contextuality in a system. These dummy random variables
are deterministic ones, attaining a single value with probability
1. The replacement of absent random variables with deterministic ones,
however, can only be made if one allows for inconsistently-connected
systems.

KEYWORDS: contextuality, dummy measurements, inconsistent connectedness,
random variables.
\end{abstract}
\maketitle
Replacing ``nothing'' with ``something'' chosen for its special
properties is one of the main ways a mathematical theory develops.
One speaks of ``nothing'' when one chooses no elements from a set,
adds no number to a total, or leaves a function unchanged; but a more
sophisticated way of speaking of these ``nothings'' would be to
take an empty subset of the set, to add a zero to the total, and to
apply an identity operator to the function. As a rule, these ``somethings''
provide not only greater convenience, but also a greater insight.
Mature set theory cannot be constructed without empty sets, nor can
algebra be developed without neutral elements of operations. One faces
an analogous situation in the theory of contextuality: ``nothing''
here means that certain things are not measured in certain contexts,
and the ``special somethings'' to replace these ``nothings'' are
deterministic random variables. 

Contextuality analysis applies to systems of random variables $R_{q}^{c}$
representing the outcomes of measuring a \emph{content} $q$ (property,
object, thing, question, sensory stimulus) in a \emph{context} $c$
(circumstances, conditions, setup). An example is the matrix below,
with three contents and four contexts:
\begin{center}
\begin{tabular}{|c|c|c|c}
\cline{1-3} 
$R_{1}^{1}$$\ensuremath{}$ & $R_{2}^{1}$$\ensuremath{}$ & $\cdot$$\ensuremath{}$ & $c=1$\tabularnewline
\cline{1-3} 
$R_{1}^{2}$$\ensuremath{}$ & $R_{2}^{2}$$\ensuremath{}$ & $\cdot$$\ensuremath{}$ & $c=2$\tabularnewline
\cline{1-3} 
$R_{1}^{3}$$\ensuremath{}$ & $\cdot$$\ensuremath{}$ & $R_{3}^{3}$$\ensuremath{}$ & $c=3$\tabularnewline
\cline{1-3} 
$\cdot$$\ensuremath{}$ & $R_{2}^{4}$$\ensuremath{}$ & $R_{3}^{4}$$\ensuremath{}$ & $c=4$\tabularnewline
\cline{1-3} 
\multicolumn{1}{c}{$q=1$} & \multicolumn{1}{c}{$q=2$} & \multicolumn{1}{c}{$q=3$} & $\boxed{\R}$\tabularnewline
\end{tabular}.
\par\end{center}

\noindent The rules such a matrix obeys are: (i) all random variables
in the same column have the same set of values (and sigma-algebras);
(ii) all random variables within a row are \emph{jointly distributed};
(iii) random variables in different rows are not jointly distributed
(are \emph{stochastically unrelated} to each other) \cite{Conversations,DzhafarovKujalaLarsson(2015),DzhafarovKujala(2016)Context-Content}.
The system is considered \emph{noncontextual} if the joint distributions
of the random variables within the rows are compatible with the joint
distributions imposed on the random variable within each column (the
compatibility meaning that both the observed row-wise distributions
and the imposed column-wise ones can be viewed as marginals of a single
probability distribution imposed on the entire system). Otherwise
the system is \emph{contextual}.

We will use the system $\R$ throughout to illustrate our points,
but the three points we make below hold for all systems of random
variables indexed by contents and contexts.

As we see in the matrix, not every content is measured in every context,
there are cells with ``nothing'' in them. It is natural to posit,
however, that for a random variable being undefined is logically equivalent
to being defined as always attaining a value labeled ``undefined.''
If so, we can fill in the empty cells with deterministic random variables, 
\begin{center}
\begin{tabular}{|c|c|c|c}
\cline{1-3} 
$R_{1}^{1}$$\ensuremath{}$ & $R_{2}^{1}$$\ensuremath{}$ & $\ensuremath{U_{3}^{1}\equiv u}$ & $c=1$\tabularnewline
\cline{1-3} 
$R_{1}^{2}$$\ensuremath{}$ & $R_{2}^{2}$$\ensuremath{}$ & $\ensuremath{\ensuremath{U_{3}^{2}\equiv u}}$ & $c=2$\tabularnewline
\cline{1-3} 
$R_{1}^{3}$$\ensuremath{}$ & $\ensuremath{U_{2}^{3}\equiv u}$ & $R_{3}^{3}$$\ensuremath{}$ & $c=3$\tabularnewline
\cline{1-3} 
$\ensuremath{U_{1}^{4}\equiv u}$ & $R_{2}^{4}$$\ensuremath{}$ & $R_{3}^{4}$$\ensuremath{}$ & $c=4$\tabularnewline
\cline{1-3} 
\multicolumn{1}{c}{$q=1$} & \multicolumn{1}{c}{$q=2$} & \multicolumn{1}{c}{$q=3$} & $\boxed{\R'}$\tabularnewline
\end{tabular},
\par\end{center}

\noindent where $u$ is interpreted as ``undefined,'' and $U\equiv u$
means that random variable $U$ equals $u$ with probability 1. In
order to comply with the rule (i) above, this value $u$ then should
be added to the set of possible values of all other random variables,
as attained by each of them with probability zero.

\emph{The first point of this note} is that a well-designed contextuality
theory should allow the addition of these deterministic $U$s to any
system without changing whether the system is\emph{ }contextual or
noncontextual. One can even implement the addition of the deterministic
$U$s empirically, e.g., by setting the procedure/device measuring
$q=3$ in contexts $c=3$ and $c=4$ to produce a fixed outcome interpreted
as ``undefined'' in contexts $c=1$ and $c=2$. 

\emph{The second point of this note} is that this desideratum cannot
be satisfied if one confines contextuality analysis to consistently-connected
systems only, the systems in which all measurements of the same content
(e.g., $R_{1}^{1}$, $R_{1}^{2}$, and $R_{1}^{3}$ in $\R$) have
the same distribution \cite{DzhafarovKujalaLarsson(2015)}. With the
exception of the Contextuality-By-Default theory, discussed below,
and of Khrennikov's conditionalization approach \cite{AvisFisherHilbertKhrennikov2009,Khrennikov2015},
this constraint is common in studies of quantum contextuality \cite{AbramskyBarbosaKishidaLalMansfield(2015),AbramskyBrandenburger(2011),Cabello(2013),KurzynskiRamanathanKaszlikowski(2012),LiangSpekkensWiseman}
(see Refs. \cite{DzhafarovKujala(2016)Fortschritte,DzhafarovKujala(2017)LNCS,KujalaDzhafarovLar(2015),DK2014Scripta}
for detailed discussions). Thus, if $R_{1}^{1}$, $R_{1}^{2}$, and
$R_{1}^{3}$ in $\R$ do not have one and the same distribution (i.e.
the system is inconsistently-connected), then, from the traditional
point of view, either the notion of contextuality is not applicable
to $\R$, or the system is considered contextual ``automatically.''
In Refs. \cite{Conversations,DzhafarovKujala(2016)Context-Content,DzhafarovKujala(2016)Fortschritte,DzhafarovKujalaLarsson(2015)}
we provide several arguments against the necessity and desirability
of the consistent connectedness constraint, and the present note adds
one more. Namely, if one agrees that the transition from $\R$ to
$\R'$ is a mere relabeling, one should consider it a flaw that in
the traditional understanding of contextuality this transition has
dramatic consequences: by adding the deterministic $U$'s to a consistently-connected
and noncontextual $\R$, one would ``automatically'' render it contextual,
or else unanalyzable in contextuality terms.

\emph{The third point of this note} is that the desideratum in question
is satisfied in the \emph{Contextuality-By-Default} (CbD) theory \cite{DzhafarovKujala(2016)Fortschritte,DzhafarovKujala(2017)LNCS,DzhafarovKujalaLarsson(2015),Conversations}:
adding the deterministic $U$s to $\R$ does not change the \emph{degree
of contextuality} computed in accordance with CbD. Moreover, the fixed
value $u$ in $\R'$ can be replaced with any other fixed values,
and different fixed values can be chosen in different cells:
\begin{center}
\begin{tabular}{|c|c|c|c}
\cline{1-3} 
$R_{1}^{1}$$\ensuremath{}$ & $R_{2}^{1}$$\ensuremath{}$ & $\ensuremath{Z_{3}^{1}\equiv z_{3}^{1}}$ & $c=1$\tabularnewline
\cline{1-3} 
$R_{1}^{2}$$\ensuremath{}$ & $R_{2}^{2}$$\ensuremath{}$ & $\ensuremath{\ensuremath{Z_{3}^{2}\equiv z_{3}^{2}}}$ & $c=2$\tabularnewline
\cline{1-3} 
$R_{1}^{3}$$\ensuremath{}$ & $\ensuremath{Z_{2}^{3}\equiv z_{2}^{3}}$ & $R_{3}^{3}$$\ensuremath{}$ & $c=3$\tabularnewline
\cline{1-3} 
$\ensuremath{Z_{1}^{4}\equiv z_{1}^{4}}$ & $R_{2}^{4}$$\ensuremath{}$ & $R_{3}^{4}$$\ensuremath{}$ & $c=4$\tabularnewline
\cline{1-3} 
\multicolumn{1}{c}{$q=1$} & \multicolumn{1}{c}{$q=2$} & \multicolumn{1}{c}{$q=3$} & $\boxed{\R^{*}}$\tabularnewline
\end{tabular}
\par\end{center}

\noindent Since the choice is arbitrary, one can always avoid the
necessity of adding, with zero probabilities, the values $z_{q}^{c}$
to the set of possible values of all $R_{q}^{c'}$, in the same column.
One can instead choose $z_{q}^{c}$ to be one of these possible values
(no matter which). Let, e.g., $R_{3}^{3}$ (hence also $R_{3}^{4}$)
in $\R$ be a binary random variable with values +1/-1; then, $Z_{3}^{1}$
can be chosen either as $Z_{3}^{1}\equiv1$ or $Z_{3}^{1}\equiv-1$.

The rest of the note demonstrates our third point. (Non)contextuality
of the system $\R$ in the CbD theory is understood as follows. \\

(A) First we introduce a certain \emph{statement} $\C$ that can be
formulated for any pair of jointly distributed random variables. This
statement should be chosen so that, for any column in $\R$, say,
$\left\{ R_{1}^{1},R_{1}^{2},R_{1}^{3}\right\} $ for $q=1$, there
is one and only one set of corresponding and jointly distributed random
variables, $\left(T_{1}^{1},T_{1}^{2},T_{1}^{3}\right)$, such that
(1) each of the $T$s is distributed as the corresponding $R$; and
(2) any two of the $T$s in $\left(T_{1}^{1},T_{1}^{2},T_{1}^{3}\right)$
satisfy the statement $\C$. This unique triple $\left(T_{1}^{1},T_{1}^{2},T_{1}^{3}\right)$
is called the $\C$-\emph{coupling} of $\left\{ R_{1}^{1},R_{1}^{2},R_{1}^{3}\right\} $,
and the $\C$-couplings for other columns of $\R$ are defined analogously.
Note that any part of the $\C$-coupling of a set of random variables
is the unique $\C$-coupling of the corresponding subset of these
random variables. In CbD, assuming all random variables in $\R$ are
binary, the role of $\C$ is played by the statement ``the two random
variables are equal to each other with maximal possible probability.''
If the measurements are not dichotomous, then the system has to be
dichotomized, as detailed in Ref. \cite{DCK2017}. We need not go
into these details, however, because we can make our point on a higher
level of abstraction, for any $\C$ with the just stipulated properties.

(B) The system $\R$ is considered $\C$-\emph{noncontextual} if there
is a random variable (vector) $S$ with jointly distributed components
corresponding to the components of $\R$,
\begin{center}
\begin{tabular}{|c|c|c|c}
\cline{1-3} 
$S_{1}^{1}$$\ensuremath{}$ & $S_{2}^{1}$$\ensuremath{}$ & $\cdot$$\ensuremath{}$ & $c=1$\tabularnewline
\cline{1-3} 
$S_{1}^{2}$$\ensuremath{}$ & $S_{2}^{2}$$\ensuremath{}$ & $\cdot$$\ensuremath{}$ & $c=2$\tabularnewline
\cline{1-3} 
$S_{1}^{3}$$\ensuremath{}$ & $\cdot$$\ensuremath{}$ & $S_{3}^{3}$$\ensuremath{}$ & $c=3$\tabularnewline
\cline{1-3} 
$\cdot$$\ensuremath{}$ & $S_{2}^{4}$$\ensuremath{}$ & $S_{3}^{4}$$\ensuremath{}$ & $c=4$\tabularnewline
\cline{1-3} 
\multicolumn{1}{c}{$q=1$} & \multicolumn{1}{c}{$q=2$} & \multicolumn{1}{c}{$q=3$} & $S$\tabularnewline
\end{tabular},
\par\end{center}

\noindent such that its rows are distributed as the corresponding
rows of $\R$ and its columns are distributed as the $\C$-couplings
of the corresponding columns of $\R$. Otherwise, if such an $S$
does not exist, the system is $\C$-\emph{contextual}. The intuition
behind this definition is that the system is $\C$-contextual if the
distributions of the random variables within contexts prevent the
random variables measuring one and the same content in different contexts
from being coupled in compliance with $\C$.

(C) If the system $\R$ is $\C$-contextual, the degree of its contextuality
is computed in the following way. The random variable $S$ above is
characterized by the probability masses 
\[
p\left(s_{1}^{1},s_{2}^{1},s_{1}^{2},s_{2}^{2},s_{1}^{3},s_{3}^{3},s_{2}^{4},s_{3}^{4}\right)
\]
assigned to every value $\left(S_{1}^{1}=s_{1}^{1},S_{2}^{1}=s_{2}^{1},\ldots,S_{3}^{4}=s_{3}^{4}\right)$
of $S$. We redefine $S$ into a \emph{quasi-random variable} if we
replace these probability masses with arbitrary real numbers 
\[
q\left(s_{1}^{1},s_{2}^{1},s_{1}^{2},s_{2}^{2},s_{1}^{3},s_{3}^{3},s_{2}^{4},s_{3}^{4}\right)
\]
summing to 1. We require that this \emph{quasi-probability distribution}
satisfy the same properties as the distribution of $S$ in (B), namely,
that it agrees with the distributions of the rows of $\R$ and with
the distributions of the $\C$-couplings of its columns. Thus, the
agreement with the first row distribution means that, for any $R_{1}^{1}=r_{1}^{1},R_{2}^{1}=r_{2}^{1}$,
we should have
\begin{equation}
\begin{array}{r}
\sum_{s_{1}^{2},s_{1}^{2},s_{1}^{3},s_{3}^{3},s_{2}^{4},s_{3}^{4}}q\left(r_{1}^{1},r_{2}^{1},s_{1}^{2},s_{2}^{2},s_{1}^{3},s_{3}^{3},s_{2}^{4},s_{3}^{4}\right)\\
\\
=\Pr\left[R_{1}^{1}=r_{1}^{1},R_{2}^{1}=r_{2}^{1}\right].
\end{array}\label{eq: bunch prob}
\end{equation}
The agreement with the distribution of the $\C$-coupling $\left(T_{1}^{1},T_{1}^{2},T_{1}^{3}\right)$
for the first column means that, for any $R_{1}^{1}=r_{1}^{1},R_{1}^{2}=r_{1}^{2},R_{1}^{3}=r_{1}^{3}$,
we should have
\begin{equation}
\begin{array}{r}
\sum_{s_{2}^{1},s_{1}^{2},s_{3}^{3},s_{2}^{4},s_{3}^{4}}q\left(r_{1}^{1},s_{2}^{1},r_{1}^{2},s_{2}^{2},r_{1}^{3},s_{3}^{3},s_{2}^{4},s_{3}^{4}\right)\\
\\
=\Pr\left[T_{1}^{1}=r_{1}^{1},T_{1}^{2}=r_{1}^{2},T_{1}^{3}=r_{1}^{3}\right].
\end{array}\label{eq: connection prob}
\end{equation}
Such quasi-random variables $S$ always exist, and among them one
can always find (generally non-uniquely) ones whose total variation
is minimal \cite{DzhafarovKujala(2016)Context-Content}. The total
variation is defined as
\begin{equation}
V\left[S\right]=\sum_{s_{1}^{1},s_{2}^{1},s_{1}^{2},s_{1}^{2},s_{1}^{3},s_{3}^{3},s_{2}^{4},s_{3}^{4}}\left|q\left(s_{1}^{1},s_{2}^{1},s_{1}^{2},s_{2}^{2},s_{1}^{3},s_{3}^{3},s_{2}^{4},s_{3}^{4}\right)\right|.\label{eq: TV}
\end{equation}
The quantity $\min V\left[S\right]-1$ can be taken as a principled
and universal measure of the\emph{ degree of contextuality}. If this
quantity equals 0, which is the smallest possible value for $V\left[S\right]-1$,
then all quasi-probability masses $q$ are nonnegative, and $S^{*}$
is a proper random variable. The system then is $\C$-noncontextual.\\

It is easy now to see the truth of our claim, that $\R^{*}$ has the
same degree of contextuality as $\R$. On the right-hand side of (\ref{eq: bunch prob}),
\[
\begin{array}{l}
\Pr\left[R_{1}^{1}=r_{1}^{1},R_{2}^{1}=r_{2}^{1}\right]\\
\\
=\Pr\left[R_{1}^{1}=r_{1}^{1},R_{2}^{1}=r_{2}^{1},Z_{3}^{1}=z_{3}^{1}\right],
\end{array}
\]
because $Z_{3}^{1}\equiv z_{3}^{1}$. The same reasoning applies to
other rows of $\R^{*}$. On the the right-hand side of (\ref{eq: connection prob}),
for any $\dot{Z}_{1}^{4}\equiv z_{1}^{4}$, 
\[
\begin{array}{l}
\Pr\left[T_{1}^{1}=r_{1}^{1},T_{1}^{2}=r_{1}^{2},T_{1}^{3}=r_{1}^{3}\right]\\
\\
=\Pr\left[T_{1}^{1}=r_{1}^{1},T_{1}^{2}=r_{1}^{2},T_{1}^{3}=r_{1}^{3},\dot{Z}_{1}^{4}=z_{1}^{4}\right].
\end{array}
\]
Now, $\left(T_{1}^{1},T_{1}^{2},T_{1}^{3},\dot{Z}_{1}^{4}\right)$
is the $\C$-coupling of $\left\{ R_{1}^{1},R_{1}^{2},R_{1}^{3},Z_{1}^{4}\right\} $.
Indeed, the $\C$-coupling $\left(\dot{T}_{1}^{1},\dot{T}_{1}^{2},\dot{T}_{1}^{3},\dot{Z}_{1}^{4}\right)$
of $\left\{ R_{1}^{1},R_{1}^{2},R_{1}^{3},Z_{1}^{4}\right\} $ exists
and is unique. The part $\left(\dot{T}_{1}^{1},\dot{T}_{1}^{2},\dot{T}_{1}^{3}\right)$
is then the unique $\C$-coupling of $\left\{ R_{1}^{1},R_{1}^{2},R_{1}^{3}\right\} $,
whence $\left(\dot{T}_{1}^{1},\dot{T}_{1}^{2},\dot{T}_{1}^{3}\right)=\left(T_{1}^{1},T_{1}^{2},T_{1}^{3}\right)$.
The same reasoning applies to other columns of $\R^{*}$. So the right-hand
sides in the equations exemplified by (\ref{eq: bunch prob}) and
(\ref{eq: connection prob}) do not change when $\R$ is replaced
with $\R^{*}$. Since, under this replacement, the left-hand sides
of these equations do not change either, except that each quasi-probability
value 
\[
q\left(s_{1}^{1},s_{2}^{1},s_{1}^{2},s_{2}^{2},s_{1}^{3},s_{3}^{3},s_{2}^{4},s_{3}^{4}\right)
\]
in them is bijectively renamed into 
\[
q\left(s_{1}^{1},s_{2}^{1},\,z_{3}^{1}\,,s_{1}^{2},s_{2}^{2},\,z_{3}^{2}\,,s_{1}^{3},\,z_{2}^{3},\,s_{3}^{3},\,z_{1}^{4},\,s_{2}^{4},s_{3}^{4}\right),
\]
the set of the quasi-probability distributions solving (\ref{eq: bunch prob})
and (\ref{eq: connection prob}) (and similar equations) in $\R^{*}$
remains the same as in $\R$, and the minimum value of $V\left[S\right]$
in (\ref{eq: TV}) therefore remains unchanged.\\

\paragraph*{Acknowledgments.}

This research has been supported by AFOSR grant FA9550-14-1-0318.
The author thanks Janne V. Kujala and Victor H. Cervantes for valuable
critical suggestions.

\end{document}